\documentclass[aps,preprint]{revtex4}
\usepackage{graphicx}

\begin{document}

\title{Atomic interaction effects in the superradiant 
light scattering from a Bose-Einstein condensate}
\author{N.\ Piovella, L.\ Salasnich, R. Bonifacio}
\affiliation{Dipartimento di Fisica, Universit\`a Degli Studi di Milano and INFM,
Via Celoria 16, Milano 20133, Italy}
\author{G.R.M.\ Robb}
\affiliation{Department of Physics, University of Strathclyde
Glasgow, G4 0NG, Scotland.}

\begin{abstract}
We investigate the effects of the atomic interaction 
in the Superradiant Rayleigh scattering from a 
Bose-Einstein condensate driven by a far-detuned laser beam. 
We show that for a homogeneous atomic sample the atomic 
interaction has only a dispersive effect, 
whereas in the inhomogeneous case it may increase the 
decay of the matter-wave grating.
\end{abstract}
\maketitle

\section{Introduction}

The long coherence time of the Bose-Einstein condensate (BEC), 
now routinely produced in many laboratories,
offers the possibility to study the collective motion induced 
by external radiation beams \cite{general}. 
In particular, a single far-off resonance laser sent on an 
elongated BEC produces superradiant Rayleigh 
scattering \cite{inouye99,kozuma99,moving}, 
generating coherent backscattered radiation and 
splitting the condensate into fractions moving at velocities 
differing by multiples of $2\hbar k/m$, 
where $k=\omega/c$ is the wave-vector of the laser incident along 
the symmetry axis of the condensate, $\omega$ is the laser frequency and $m$ is the atomic mass. 
Superradiant Rayleigh scattering from a BEC is the 
quantum analog of the collective atomic recoil laser (CARL) \cite{CARL} in which
the emitted radiation is not confined in a high-$Q$ ring cavity \cite{CARL-SR}. 
The complete absence in a BEC of Doppler broadening due 
to thermal motion allows for a regime in which the atoms scatter a single laser photon and recoil with an
extra momentum of $2\hbar k$ in the direction of the incident photon. The number of scattered photons and the
amplitude of the density grating resulting from the interference between the two atomic wavepackets with momentum 
difference $2\hbar k$ are exponentially enhanced via the CARL instability \cite{moore}. 
In the absence of any mechanism of atomic dephasing, the process is sequential 
\cite{inouye99,gatelli}, with a complete transfer 
of atoms, after the superradiant process, from the original motional state with 
momentum $p$ to a state with a momentum 
$p+2\hbar k$. However, in a real BEC several mechanisms contribute to the 
decay of the coherence between the 
two momentum states. Some of them are due to the decoherence induced
by spontaneous emission \cite{spont} or phase diffusion \cite{moving}. 
The main characteristic of this kind of decoherence is irreversibility. 
Other mechanisms arise from inhomogeneous broadening, as those due to a
finite size of the condensate wavefunction, responsible for a broadening of the atomic momentum distribution, 
and from mean-field broadening due to the atomic interaction \cite{stenger99}.
It has been recently suggested that the dephasing due to inhomogeneous broadening can be reversed 
applying, after the superradiant scattering process, a Bragg pulse of area $\pi$  inducing a superradiant 
echo and a further tranfer of the atoms to the final momentum state \cite{echo}. 
In this paper we investigate the effects of the mean-field atomic interaction on the superradiant scattering process.

\section{Basic model}
  
We consider an elongated Bose-Einstein condensate driven by a single
laser incident along the positive direction of the symmetry axis $z$ of 
the condensate. The laser is far-detuned from the atomic resonance, so that radiation
pressure due to absorption and subsequent random incoherent, isotropic
emission of a photon, can be neglected. In this regime, the atoms backscatter photons at frequency $\omega_s$
and wave vector $k_s=\omega_s/c\approx k$, recoiling with a momentum $2\hbar k$ along the same direction of the 
incident laser beam.

In a simplified 1D description of the process along the axis $z$, 
the evolution of the matter-wave field $\Psi(z,t)$ and of the dimensionless amplitude $a(t)$ of the scattered radiation
is determined by the following self-consistent equations:
\begin{eqnarray}
i\frac{\partial\Psi}{\partial
t}&=&-\omega_r
\frac{\partial^2\Psi}{\partial\theta^2}+ig
\left[a^* e^{i(\theta-\delta t)}- {\rm c.c.}\right]\Psi+2\pi\beta |\Psi|^2\Psi
\label{psi}\\
\frac{da}{dt}&=&gN \int d\theta|\Psi|^2e^{i(\theta-\delta t)}-\kappa a.
\label{a}
\end{eqnarray}
where  $\theta=2kz$, $a=(\epsilon_0 V/2\hbar\omega_s)^{1/2}E$ is the
dimensionless electric field amplitude of the scattered beam with
frequency $\omega_s$, $\omega_r=2\hbar k^2/m$ is the two-photon recoil 
frequency, 
$g=(\Omega/2\Delta_0)(\omega d^2/2\hbar\epsilon_0 V)^{1/2}$ is the 
coupling constant, 
$\Omega$ is the Rabi frequency of the laser beam of frequency $\omega=ck$, 
detuned from the atomic resonance frequency $\omega_0$ by $\Delta_0=\omega-\omega_0$, 
$d$ is the electric dipole moment of the atom along the laser polarization direction,
$V$ is the volume of the condensate containing $N$ atoms, 
$\delta=\omega-\omega_s$ and $\epsilon_0$ is the permittivity of the free space.
The second term on the right hand side of Eq.(\ref{psi}) is the self-consistent 
optical lattice, resulting from the interference between the laser and the backscattered radiation, 
whose amplitude is amplified by the matter-wave grating described by the first term on the right hand 
side of Eq.(\ref{a}).
The matter-wave field $\Psi$ is normalized to one, i.e. $\int d\theta|\Psi(\theta,t)|^2=1$, and
the last term on the right hand side of Eq.(\ref{psi}) describes the atomic 
interaction due to binary collisions, where $\beta=4\hbar k a_sN/m\Sigma$,
$a_s$ is the scattering length and $\Sigma$ is the condensate cross section. 
Eq.(\ref{a}) has been written in the ``mean-field''
limit, which models the propagation effects of the light by
replacing the nonuniform electric field by an average value and by
adding to the equation a damping term with decay constant
$\kappa\approx c/2L$, where $L$ is the condensate length and $c$
is the speed of light in vacuum.

\section{Homogeneous case}

If the condensate is much longer than the radiation wavelength and the density is uniform, then
periodic boundary conditions can be assumed on $\theta$ and the wavefunction can be 
written as a Fourier series
\begin{equation}
\Psi(\theta,t)=\sum_n c_n(t)u_n(\theta)e^{-in\delta t},\label{pbc}
\end{equation}
where $u_n(\theta)=(1/\sqrt{2\pi})\exp(in\theta)$ are momentum eigenfunctions 
with eigenvalues $p_z=n(2\hbar k)$. Using Eq.(\ref{pbc}), Eqs.(\ref{psi}) 
and (\ref{a}) reduce to an 
infinite set of ordinary differential equations,
\begin{eqnarray}
\dot c_n&=& -i\delta_nc_n+g(a^*c_{n-1}-ac_{n+1})
-i\beta\sum_{m,l}c_m c_l c^*_{m+l-n}\label{cn}\\
\dot a&=& g N \sum_{n}c_n c^*_{n+1}-\kappa a\label {a:2},
\end{eqnarray}
where $\delta_n=n^2\omega_r-n\delta$ and the dot indicates 
the time derivative.

\subsection{Two-level approximation}

Assuming that the only two momentum levels involved in the process are the 
initial level $n$ and the final level $n+1$, Eq.(\ref{cn}) and (\ref{a:2}) reduce to:
\begin{eqnarray}
\dot c_n&\approx& -i\left[\delta_n+\beta\left(|c_n|^2+2|c_{n+1}|^2\right)
\right]c_n
-gac_{n+1} \label{c1}\\
\dot c_{n+1}&\approx&-i\left[\delta_{n+1}+\beta\left(2|c_n|^2+|c_{n+1}|^2
\right)\right]c_{n+1}
+ga^*c_n\label{c2}\\
\dot a &\approx& g N c_n c^*_{n+1}-\kappa a\label {a:3}.
\end{eqnarray}
Defining $S=c_nc^*_{n+1}$ and $W=|c_n|^2-|c_{n+1}|^2$, we obtain from 
Eqs.(\ref{c1})-(\ref{a:3}):
\begin{eqnarray}
\dot S&=& -i\left(\Delta-\beta W\right)S+gaW
-\gamma S\label{S}\\
\dot W&=& -2g\left(aS^* + {\rm c.c.}\right)\label{W}\\
\dot a &=& g N S-\kappa a\label {a:4},
\end{eqnarray}
where $\Delta=\delta_n-\delta_{n+1}=\delta-\omega_r(2n+1)$ and
we have introduced a damping term in Eq.(\ref{S}), to account for the 
decay of the coherence between the two motional states $n$ and $n+1$. 
We note that when the atomic interaction is neglected ($\beta=0$), $\Delta=0$ is the Bragg condition of 
the scattering process, arising from momentum and energy conservation \cite{martin88}.
We observe from Eq.(\ref{S}) that the atomic interaction term has a dynamical dispersive effect on the Bragg 
resonance, proportional to the population difference $W$. 
In the linear regime, when $a$ is still small and $W\approx 1$, the Bragg  
condition is $\Delta=\beta$, i.e. $\delta=\omega_r(2n+1)+n_aU/\hbar$, 
where $n_a U=n_a4\pi\hbar^2a_s/m$ is the chemical 
potential and $n_a=2N/\lambda\Sigma$ is the atomic density.

\begin{figure}[t]
\includegraphics[width=8cm]{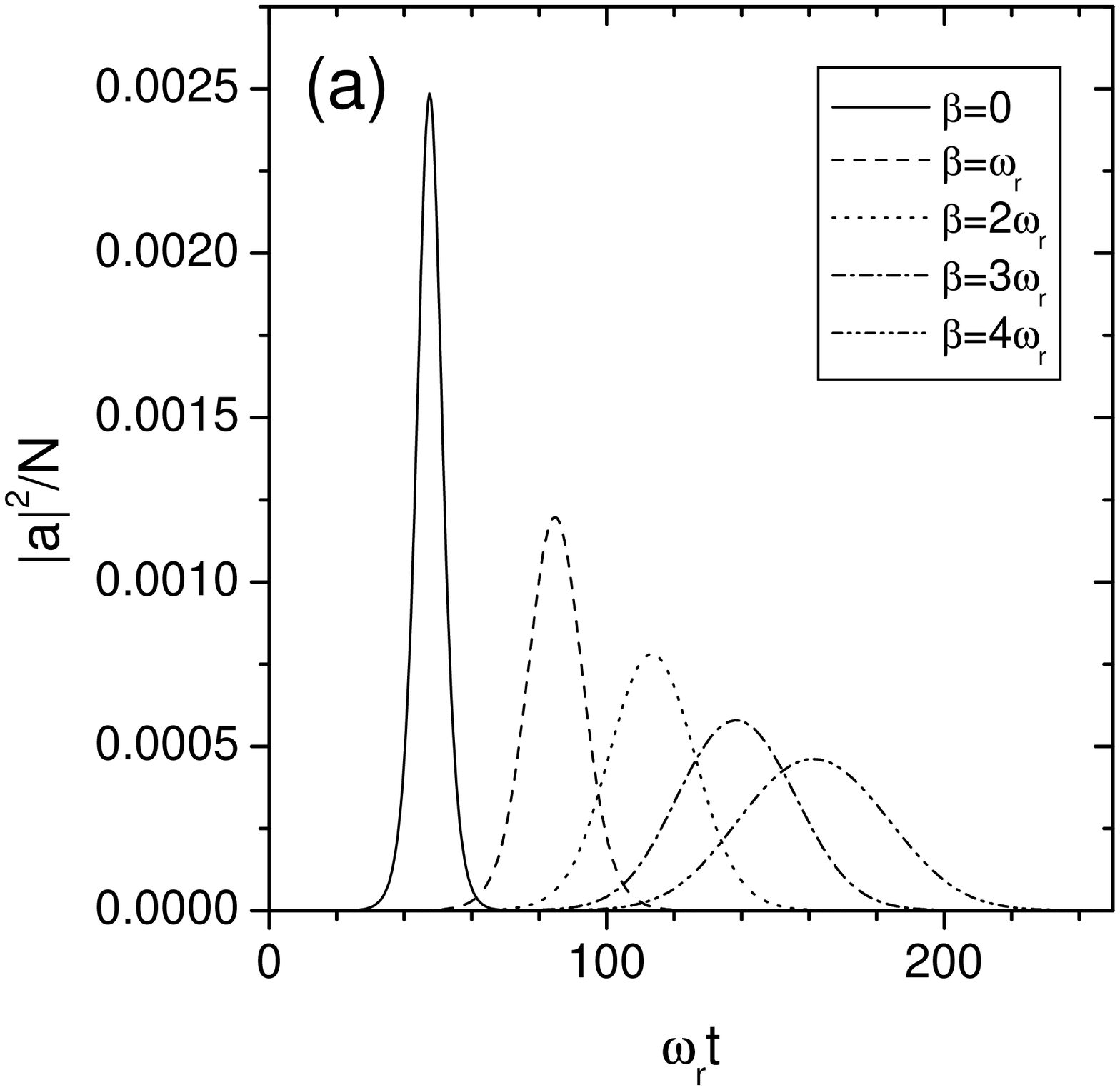}
\includegraphics[width=8cm]{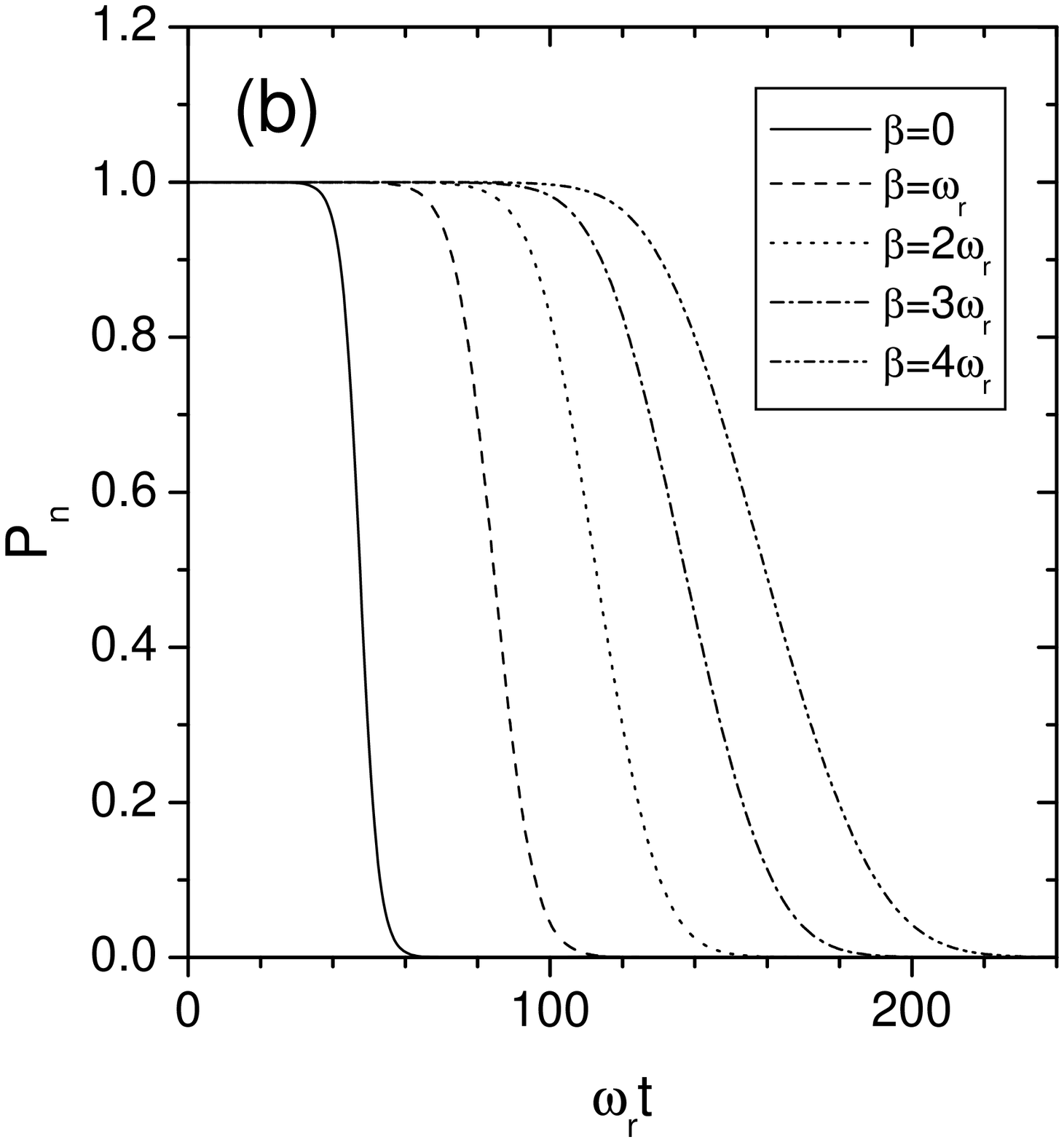}
\begin{center}
\caption{Effects of the atomic interaction on the superradiant regime 
in the homogeneous case:  $|a|^2/N$, (a), and population fraction $P_n$ of the initial momentum state, (b),
vs. $\omega_r t$, from the numerical integration of Eqs.(\ref{cn}) and (\ref{a:2}) with 
$\kappa=20\omega_r$, $g\sqrt {N}=2\omega_r$, $\Delta=\beta$ and different values of $\beta$.} 
\label{fig1}
\end{center}
\end{figure}

In the superradiant regime the field amplitude $a$ can be 
adiabatically eliminated for times much longer than 
$\kappa^{-1}$. In fact, let introduce the slowly varying variable 
$\tilde S(t)=S(t)e^{i\alpha(t)}$, where $\alpha(t)\Delta t-\beta\int_0^t dt'W(t')$, and let integrate Eq.(\ref{a:4}):
\begin{equation}
a(t)=a(0)e^{-\kappa t}+gN\int_0^t dt'
\tilde S(t-t')e^{-i\alpha(t-t')-\kappa t'}.
\label{adia1}
\end{equation}
If we assume that $\tilde S$ and $W$ do not change appreciably 
in a time $\kappa^{-1}$ during the superradiant process, i.e.
if $\tau_{sr}\gg\kappa^{-1}$ where $\tau_{sr}$ is a characteristic time for
superradiance, then in Eq.(\ref{adia1}) $\tilde S(t-t')\approx\tilde S(t)$ and 
$\alpha(t-t')\approx\alpha(t)-[\Delta-\beta W(t)] t'$. 
Performing the residual integration in $t'$ and assuming 
$t\gg\kappa^{-1}$, we finally obtain:
\begin{equation}
a(t)\approx \frac{gNS(t)}{\kappa-i[\Delta-\beta W(t)]},\label{adiab}
\end{equation}
so that the field $a$ follows istantaneously the atomic evolution. 
Combining Eqs.(\ref{S}), (\ref{W}) and (\ref{adiab}) and 
defining $I=|S|^2$, we obtain 
\begin{eqnarray}
\dot I&=& 2\left\{\frac{GW}{1+[(\Delta-\beta W)/\kappa]^2}-
\gamma\right\}I\label{I}\\
\dot W&=& -\frac{4GI}{1+[(\Delta-\beta W)/\kappa]^2}\label{W:2}
\end{eqnarray}
where $G=g^2N/\kappa$ is the superradiant gain. 
Defining the characteristic time as $\tau_{sr}=1/G$, it follows 
that the adiabatic approximation (\ref{adiab}) is true for $\kappa\gg G$, 
i.e. for $\kappa\gg g\sqrt N$. Furthermore,
the two-level approximation is valid if $G<\omega_r$.
>From Eqs.(\ref{I}) and (\ref{W:2}) it is possible to derive the following analytical results: 

\begin{itemize}
\item In the linear regime, for $W\approx 1$, the threshold 
condition for superradiance is 
$G>\gamma\{1+[(\Delta-\beta)/\kappa]^2\}$, so that the only effect 
of the atomic interaction is a shift of 
the resonance from $\Delta=0$ to $\Delta=\beta$.

\item Neglecting decoherence ($\gamma=0$), Eqs. (\ref{I}) 
and (\ref{W:2}) admit the constant of 
motion $4I+W^2=1$. Writing $2\sqrt{I}=\sin\phi$ and $W=\cos\phi$, we obtain an equation for the Bloch angle
$\phi$:
\begin{equation}
\dot\phi=\frac{G\sin\phi}{1+\left(\Delta-\beta\cos\phi\right)^2/\kappa^2}.
\label{phi}
\end{equation}
Although Eqs.(\ref{phi}) can be solved exactly by quadrature, 
its solution can not be set in an explicit form.  
Fig.\ref{fig1} shows $|a|^2/N$, (a), and $P_n=|c_n|^2=(W+1)/2$, (b), 
as a function of $\omega_r t$ for different values of $\beta$, obtained solving numerically
Eqs.(\ref{cn}) and (\ref{a:2}) for $\kappa=20\omega_r$,
$g\sqrt{N}=2\omega_r$ and $\Delta=\beta$. 
The results are in excellent agreement with the numerical solution of the approximated Eq.(\ref{phi}), 
not reported in the figure. 
We note that the effect of the atomic interaction is only a broadening of the 
superradiant pulse, which still preserves the same area
equal to $\pi$, transfering completely  the atoms from the initial 
momentum state $n$ to the final momentum state $n+1$.
\item It is easy to calculate the exact analytical solution of Eqs. (\ref{I}) 
and (\ref{W:2}) when 
$\beta=0$. In fact, let introduce $G'=G/[1+(\Delta/\kappa)^2]$ and 
the new variables
$x=(W-W_0)/(1-W_0)$ and $y=2\sqrt{I}/(1-W_0)$, where $W_0=\gamma/G'<1$. 
>From Eqs. (\ref{I}) and (\ref{W:2}) it 
follows that $x^2+y^2=1$. Introducing again the Bloch angle $\phi$ defined 
such that $x=\cos\phi$ and $y=\sin\phi$,
Eqs. (\ref{I}) and (\ref{W:2}) give the following equation for $\phi$,
\begin{equation}
\dot\phi=G'(1-W_0)\sin\phi,\label{phi:3},
\end{equation}
whose solution yields $x(t)=-\tanh[G'(1-W_0)(t-t_D)]$, where 
$t_D=-\ln[|S(0)|/(1-W_0)]/G'(1-W_0)$ is the delay time. Coming back to the original variables we finally
obtain:
\begin{eqnarray}
I(t)&=&\left(\frac{1-W_0}{2}\right){\rm sech}^2[G'(1-W_0)(t-t_D)]\label{I:3}\\
W(t)&=& W_0-(1-W_0)\tanh[G'(1-W_0)(t-t_D)]\label{W:3}
\end{eqnarray}
We note that the asymptotic value of the population difference is $2W_0-1$, 
so that the fraction of atoms left in the initial
state after the superradiant process is $P_n=W_0=\gamma/G'$. Measuring experimentally $G'$ and $P_n$ 
it is possible to evaluate the decoherence rate $\gamma$ \cite{moving}.

\end{itemize}

\section{Inhomogeneous case}

\begin{figure}
\includegraphics[width=15cm]{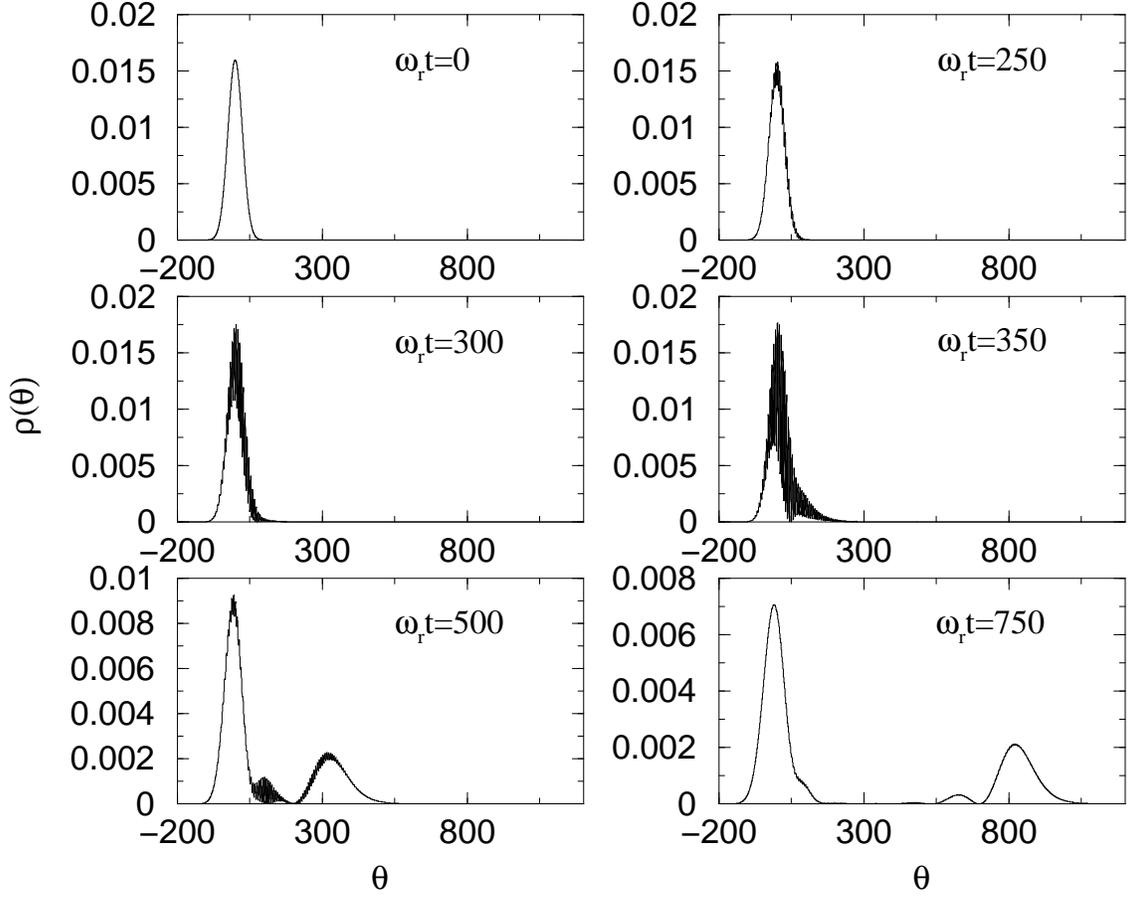}
\begin{center}
\caption{Evolution of the density distribution $\rho(\theta,t)=|\Psi(\theta,t)|^2$ vs. $\theta=2kz$
at different times, from the numerical integration of Eqs.(\ref{psi}) and (\ref{a}) for
an initial Gaussian wavepacket of width $\sigma_\theta=25$, $\kappa=10\omega_r$, 
$g\sqrt {2N}=\omega_r$, $\Delta=0$, and $\beta=0$.} 
\label{fig2}
\end{center}
\end{figure}

\begin{figure}[t]
\includegraphics[width=15cm]{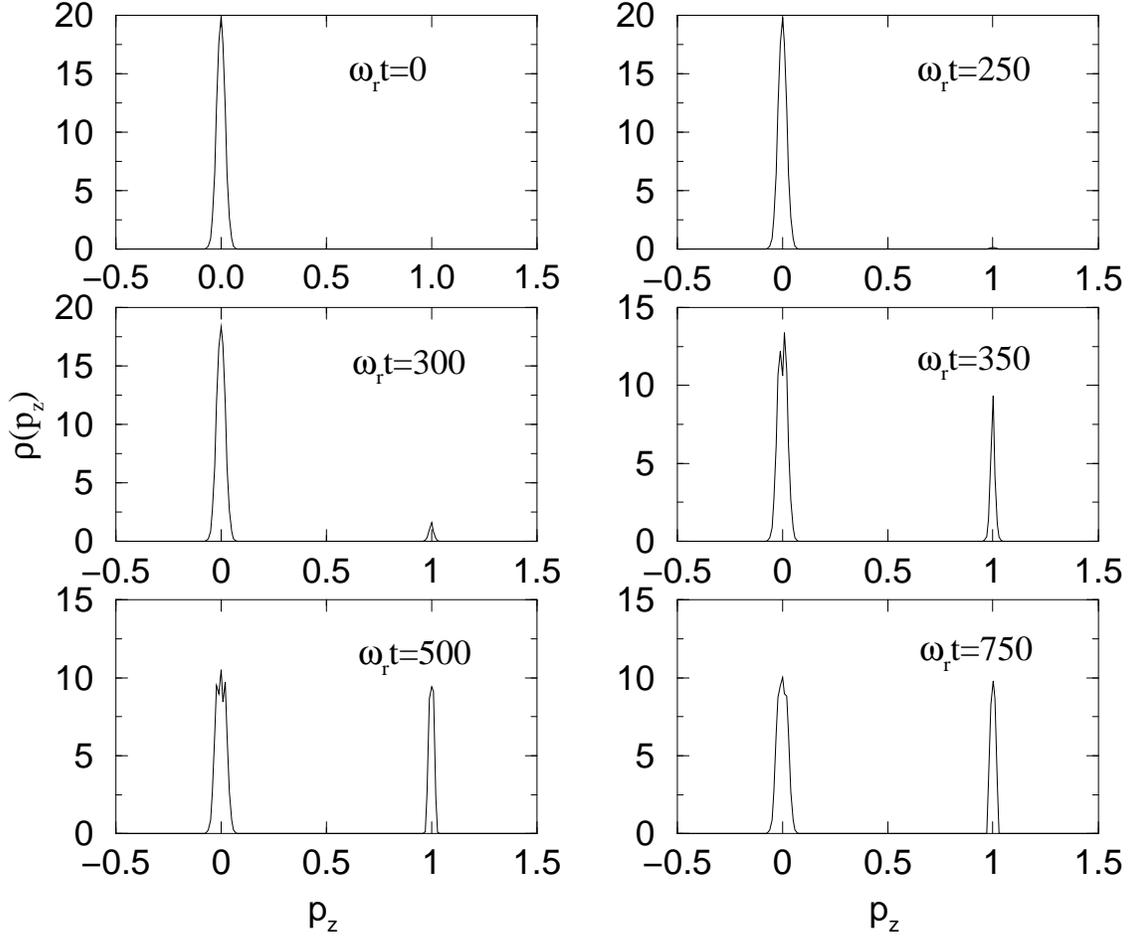}
\begin{center}
\caption{Evolution of the momentum distribution $\rho(p_z,t)=|\tilde\Psi(p_z,t)|^2$ vs. $p_z$ (in units of $2\hbar k$) 
at different times and for the same case shown in fig.\ref{fig2}.} 
\label{fig3}
\end{center}
\end{figure}

Let now consider the case in which the condensate is described initially by a 
wavepacket with a finite size $\sigma_\theta=2k\sigma_z$ and $p_z=0$. We have solved Eqs.(\ref{psi}) and (\ref{a}) 
for an initial Gaussian wavepacket of width
$\sigma_\theta=25$, $\kappa=10\omega_r$, 
$g\sqrt{2N}=\omega_r$, $\Delta=0$ and different values of $\beta$. 
The numerical integration of Eqs. (\ref{psi}) and (\ref{a}) 
is based on a finite-difference predictor-corrector 
scheme \cite{sala1,sala2}.
Fig.\ref{fig2} shows the density distribution, $\rho(\theta,t)=|\Psi(\theta,t)|^2$, for $\beta=0$ at different times,
whereas fig.\ref{fig3} shows the corresponding momentum distribution $\rho(p_z,t)=|\tilde\Psi(p_z,t)|^2$,
where $\tilde\Psi$ is the Fourier transform of the wavefunction $\Psi$.
We observe that the superradiant process produces a condensate fraction moving with an average momentum 
$p_z=2\hbar k$ and a smaller momentum spread (see fig.\ref{fig3}). 
In the configuration space (see fig.\ref{fig2}) we clearly observe the interference fringes when the two 
fractions overlap, whereas for longer times the recoiling atoms move away from the original condensate.

\begin{figure}[b]
\includegraphics[width=10cm]{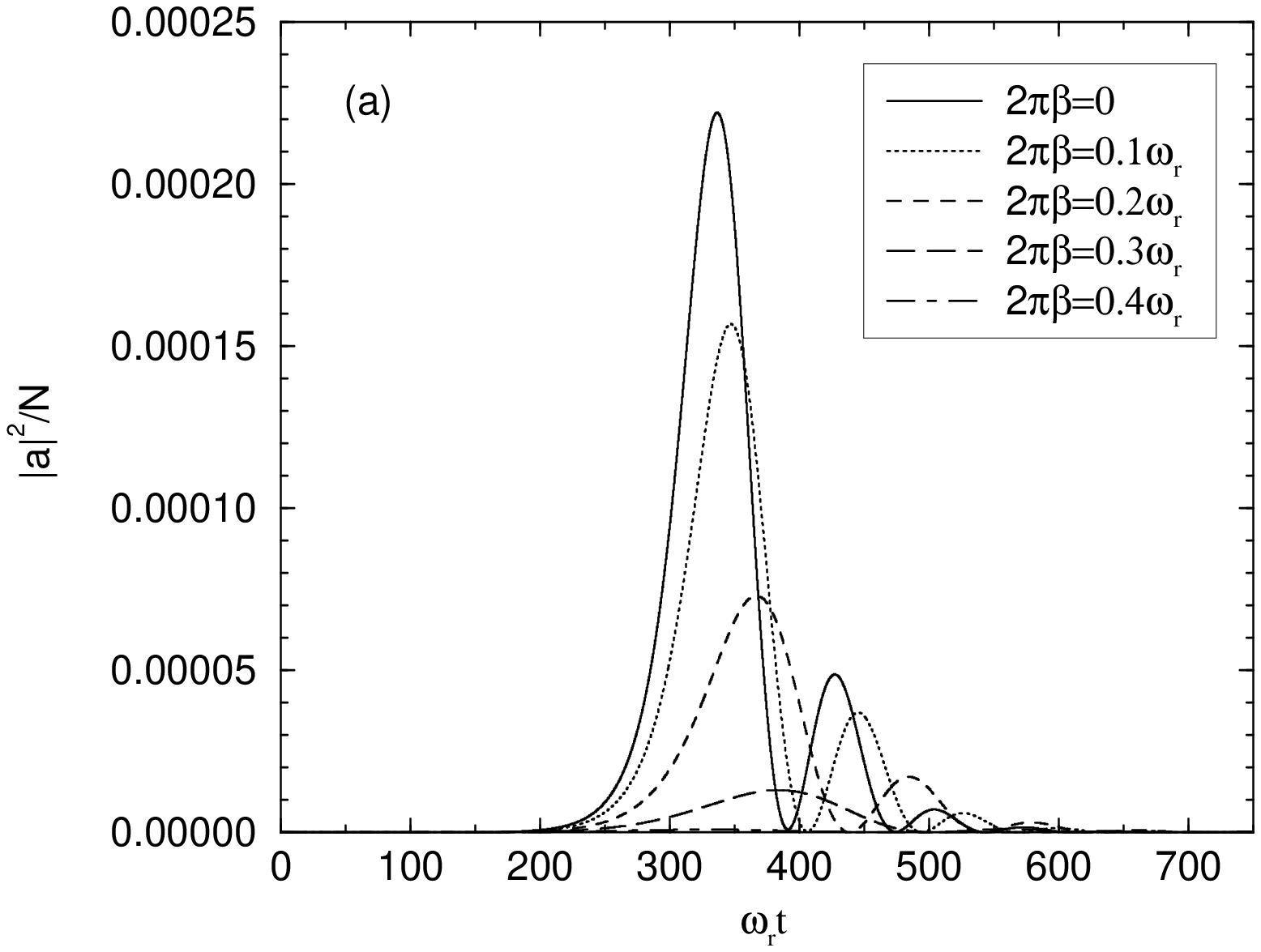}
\includegraphics[width=10cm]{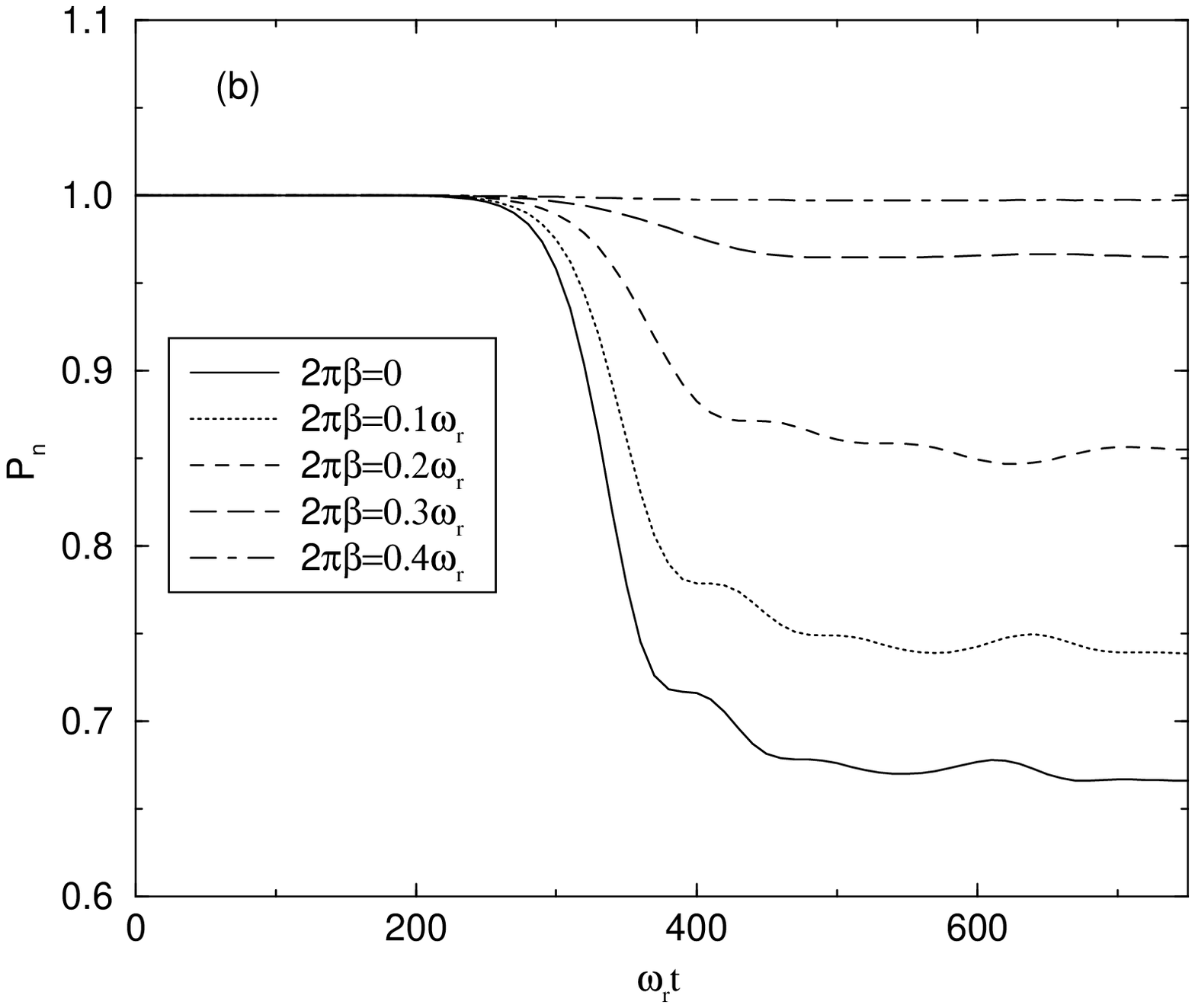}
\begin{center}
\caption{Effects of the atomic interaction on the superradiant regime 
in the inhomogeneous case:  $|a|^2/N$, (a), and population fraction $P_n$ of the initial momentum state, (b),
vs. $\omega_r t$, for the same initial conditions and parameters of the case shown in fig.\ref{fig2} and  
different values of $\beta$.} 
\label{fig4}
\end{center}
\end{figure}

Fig.\ref{fig4} shows $|a|^2/N$, (a), and $P_n$, (b), as a 
function of $\omega_r t$ for different values of $\beta$, where $P_n$ is the population of the 
momentum state $p_z=n(2\hbar k)$, calculated integrating the momentum 
distribution  over an interval centered around $p_z=n(2\hbar k)$ and of length $2\hbar k$. 
This can be done only if the momentum distribution remains narrower 
than the momentum level separation, i.e. if $\sigma_{p_z}\ll 2\hbar k$.

We observe that contrary to the homogeneous case, increasing 
$\beta$ the superradiant process becomes less efficient,
decreasing the area of the superradiant pulse (see fig.\ref{fig4}(a)) and increasing 
the fraction of atoms left in the initial momentum state (see fig.\ref{fig4}(b)). 
This effect can be interpreted as due to a dephasing caused 
by a detuning from the resonance depending on the
atomic density. Each atom evolves with a different detuning 
from the resonance, resulting in a inhomogeneous broadening of
the superradiant transition and a subsequent decay of the coherence 
between the atoms. Similarly to the photon 
echo \cite {echo}, it is expected that this dephasing may be partially 
reversed applying a suitable Bragg pulse or area
$\pi$, at least for small values of $\beta$. 
For larger or negative values of $\beta$, it is expected that nonlinear effects are more important, 
and this regime will be the object of a future detailed analysis.

\newpage

\end{document}